\documentclass[twocolumn,prb,showpacs,byrevtex,superbib]{revtex4}
\usepackage{amsmath}
\usepackage{graphicx}
\usepackage{epsfig}
\usepackage[bf]{subfigure}
\usepackage[scaled=.92]{helvet}
\usepackage{courier}

\input{tcilatex}
\DeclareMathAlphabet{\mathsf}{OT1}{cmss}{bx}{n}

\begin{document}

\title
[Rays, modes, and wavefield stability]{Rays, modes, wavefield structure and wavefield stability}%

\author{Michael G. Brown$^{\textrm{a})}$\footnotetext{$^{\textrm{a})}$Author to whom correspondence should be addressed. Electronic mail: \texttt{mbrown@rsmas.miami.edu}}}%

\affiliation{RSMAS/AMP, University of Miami, Miami, FL 33149}%

\author{Francisco J. Beron-Vera}%

\affiliation{RSMAS/AMP, University of Miami, Miami, FL 33149}%

\author{Irina Rypina}%

\affiliation{RSMAS/AMP, University of Miami, Miami, FL 33149}%

\author{Ilya Udovydchenkov}%

\affiliation{RSMAS/AMP, University of Miami, Miami, FL 33149}%



\begin{abstract}%

Sound propagation is considered in range-independent environments and
environments consisting of a range-independent background on which a weak
range-dependent perturbation is superimposed. Recent work on propagation of
both types of environment, involving both ray- and mode-based wavefield
descriptions, have focused on the importance of $\alpha $, a ray-based
``stability parameter,'' and $\beta ,$ a mode-based ``waveguide invariant.''
It is shown that, when $\beta $ is evaluated using asymptotic mode theory, $%
\beta =\alpha .$ Using both ray and mode concepts, known results relating to
the manner by which $\alpha $ (or $\beta $) controls both the unperturbed
wavefield structure and the stability of the perturbed wavefield are briefly
reviewed.

\end{abstract}%

\volumeyear{2002}%

\volumenumber{205}%

\issuenumber{5}%

\date[Dated: ]{\today}%

\startpage{1}%

\endpage{102}%

\eid{identifier}%

\maketitle%

\section{Introduction}

In recent work \cite%
{Beron-Brown-03a,Beron-Brown-03b,Virovlyansky-03,Beron-etal-03,Smirnov-etal-01,Brown-etal-03}
on sound propagation in the ray limit in weakly range-dependent ocean
environments it has been shown that both ray amplitude and phase (travel
time) distributions are largely controlled by a property---described by the 
\textit{stability parameter} $\alpha $---of the background sound speed
profile. Other investigators \cite%
{Brekhovskikh-Lysanov-91,Dspain-Kuperman-99,Chuprov-82,Grachev-93,Weston-Stevens-72,Kuzkin-95,Kuzkin-etal-98,Kuzkin-98,Kim-etal-03,Hodgkiss-etal-99,Dspain-etal-99,Song-etal-98}
making use of a modal description of the wavefield in both range-independent
environments and range-independent environments with weak perturbations
superimposed have shown that many wavefield properties are controlled by a
property---the \textit{waveguide invariant} $\beta $---which is defined
using mode-based quantities in the background environment. In this letter it
is shown that when $\beta $ is evaluated using asymptotic mode-theoretical
results, $\beta =\alpha .$

In the two sections that follow $\alpha $ and $\beta ,$ respectively, are
defined and briefly discussed. In the final section those properties of
wavefields that are known to be controlled by $\alpha $ or $\beta $ are
briefly reviewed. To keep our presentation brief we show only equations
required to derive the result $\beta \sim \alpha $ or those that provide
insight into wavefield properties discussed in the final section.

\section{Rays: $\mathbf{\protect\alpha }$}

We consider underwater acoustic wavefields in three space dimensions that
are excited by a point source. We assume that the environment consists of a
depth-dependent background, with a weak perturbation superimposed, so the
sound speed is written as $c(z,r)=C(z)+\delta c(z,r).$ Here $-z$ is depth
and the range $r$ is the horizontal distance from the source. When $\delta
c\neq 0$ it is assumed that azimuthal coupling of the wavefield is
negligible, so that one need only consider propagation in the $(r,z)$ plane.
To make our discussion concrete we will assume that $C(z)$ has a single
minimum, but this assumption is not necessary.

It is well known (cf. e.g. Refs. 
\onlinecite{Smirnov-etal-01}%
,%
\onlinecite{Brown-etal-03}%
) that substitution of the geometric ansatz%
\begin{equation}
\bar{u}(z,r,\sigma )=a(z,r)\mathrm{e}^{\mathrm{i}\sigma T(z,r)},
\end{equation}%
where $\bar{u}(z,r,\sigma )$ is the Fourier transform of the pressure $%
u(z,r,t)$ and $\sigma $ is the acoustic frequency, into the Helmholtz
equation and collecting terms in descending powers of $\sigma $ yields the
eikonal and transport equations. The eikonal equation can be solved for the
travel time $T$ by integrating the ray equations,%
\begin{equation}
\frac{\mathrm{d}p_{z}}{\mathrm{d}r}=-\frac{\partial H}{\partial z},\quad 
\frac{\mathrm{d}z}{\mathrm{d}r}=\frac{\partial H}{\partial p_{z}},\quad 
\frac{\mathrm{d}T}{\mathrm{d}r}=p_{z}\frac{\mathrm{d}z}{\mathrm{d}r}-H
\label{AA}
\end{equation}%
where%
\begin{equation}
H(p_{z},z,r)=-\sqrt{c^{-2}(z,r)-p_{z}^{2}}.
\end{equation}%
It follows from these equations and the relationship $\mathrm{d}z/\mathrm{d}%
r=\tan \varphi ,$ where $\varphi $ is ray angle with respect to the
horizontal, that $cp_{z}=\sin \varphi ,$ so the vertical slowness $p_{z}$
can be thought of as a scaled angle variable. For a point source it is
convenient to label rays by their $p_{z}$ value at $(z,r)=(z_{0},0),$ $%
p_{z,0}.$ The \textit{Hamiltonian} $H=-p_{r},$ where $p_{z}$ and $p_{r}$ are
the vertical and horizontal components of the slowness vector $\mathbf{p}$
with $\left\| \mathbf{p}\right\| =c^{-1}(z,r).$ In a range-independent
environment $c=C(z),$ $p_{r}$ is constant following a ray. The transport
equation can be reduced to a statement of the constancy of energy flux in
ray tubes; its solution, accounting for azimuthal spreading, can be written%
\begin{equation}
a^{2}=a_{0}^{2}\frac{r_{0}^{2}}{r}\left| H\left( \partial z/\partial
p_{z,0}\right) _{r}\right| ^{-1}.  \label{a2}
\end{equation}%
Here $a_{0}^{2}$ is the value of $a^{2}$ at the small distance ($1$ \textrm{%
m }by convention) $r_{0}$ from the source. The partial derivative in (\ref%
{a2}) is evaluated keeping $r$ fixed: $z(p_{z,0},r)$ is ray depth. In a
range-independent environment an alternative form of (\ref{a2}) is%
\begin{equation}
a^{2}=a_{0}^{2}\frac{r_{0}^{2}}{r}\frac{\left| p_{r}/p_{z,0}\right| }{\left|
p_{z}\left( \partial R/\partial p_{r}\right) _{z}\right| },  \label{a22}
\end{equation}%
where $R(p_{r},z)$ is the range of a ray, and ray depth is held constant in
the partial derivative.

For the range-independent problem $z(r)$ and $p_{z}(r)$ (following rays) are
periodic functions. This periodic motion is most naturally described using 
\textit{action--angle variables} $(I,\vartheta )$. The transformed ray
equations (cf. e.g. Refs. 
\onlinecite{Beron-Brown-03b}%
,%
\onlinecite{Virovlyansky-03}
for details of the transformation) are%
\begin{eqnarray}
\frac{\mathrm{d}I}{\mathrm{d}r} &=&-\frac{\partial \bar{H}}{\partial
\vartheta }=0, \\
\frac{\mathrm{d}\vartheta }{\mathrm{d}r} &=&\frac{\partial \bar{H}}{\partial
I}=\omega (I), \\
\frac{\mathrm{d}T}{\mathrm{d}r} &=&I\omega (I)-\bar{H}(I)  \label{AAeqn}
\end{eqnarray}%
where $\bar{H}(I)$ is the transformed Hamiltonian. The action 
\begin{equation}
I=\frac{1}{2\pi }\int_{z_{-}}^{z_{+}}\mathrm{d}z\,p_{z}(z)=\frac{1}{\pi }%
\int_{z_{-}}^{z_{+}}\mathrm{d}z\,\sqrt{C^{-2}(z)-p_{r}^{2}},  \label{I}
\end{equation}%
where $z_{\pm }$ correspond to the ray upper $(+)$ and lower $(-)$ turning
depths where $C^{-1}(z_{-})=C^{-1}(z_{+})=p_{r}.$ The angle variable $%
\vartheta $ increases by $2\pi $ each time a ray completes a cycle, and $%
\omega (I)=2\pi /R_{\ell }(I)$ where $R_{\ell }$ is the range of a ray cycle
(double loop),%
\begin{equation}
R_{\ell }(p_{r})=-2\pi \frac{\mathrm{d}I}{\mathrm{d}p_{r}}%
=2p_{r}\int_{z_{-}}^{z_{+}}\frac{\mathrm{d}z}{\sqrt{C^{-2}(z)-p_{r}^{2}}}
\label{R}
\end{equation}%
where $I(p_{r})$ is defined in (\ref{I}). The action--angle form of the ray
equations can trivially be integrated: 
\begin{eqnarray}
\vartheta (r) &=&\vartheta _{0}+\omega (I)r\text{ }\func{mod}2\pi \\
I(r) &=&I_{0} \\
T(r) &=&\left[ I\omega (I)-\bar{H}(I)\right] r.  \label{AAtra}
\end{eqnarray}%
[More correctly, a term $\mathrm{d}\left( G-I\vartheta \right) /\mathrm{d}r$
should be added to the r.h.s. of (\ref{AAeqn}) where $G$ is the generating
function of the canonical transformation $(p_{z},z)\rightarrow (I,\vartheta
).$ The corresponding endpoint corrections to (\ref{AAtra}) is both
numerically insignificant and not relevant to the discussion that follows.]

The stability parameter $\alpha (I),$ whose relationship to wavefield
properties will be reviewed in the final section, is defined as 
\begin{equation}
\alpha (I)=\frac{I}{\omega (I)}\frac{\mathrm{d}\omega }{\mathrm{d}I}.
\label{alpha}
\end{equation}%
It follows from the relationship $\omega (I)=2\pi /R_{\ell }(I)$ that $%
\alpha (I)$ can be expressed in the form%
\begin{equation}
\alpha (p_{r})=2\pi \frac{I(p_{r})}{R_{\ell }^{2}(I)}\frac{\mathrm{d}R_{\ell
}}{\mathrm{d}p_{r}}.  \label{alpha2}
\end{equation}

\section{Modes: $\mathbf{\protect\beta }$}

In a stratified environment $c=C(z)$ in three space dimensions the modal
decomposition of the wavefield has the form%
\begin{equation}
\bar{u}(z,r,\sigma )=\frac{\mathrm{i}}{4}\sum_{m=0}^{\infty }\frac{\phi
_{m}(z_{0},\sigma )\phi _{m}(z,\sigma )H_{0}^{(1)}(\sigma p_{rm}r)}{\int 
\mathrm{d}z\,\phi _{m}^{2}(z,\sigma )}.
\end{equation}%
Here $H_{0}^{(1)}$ is the zeroth-order Hankel function of the first kind and
the normal modes $\phi _{m}(z;\sigma )$ satisfy%
\begin{equation}
\frac{\mathrm{d}^{2}\phi _{m}}{\mathrm{d}z^{2}}+\sigma ^{2}\left[
C^{-2}(z)-p_{rm}^{2}\right] \phi _{m}=0  \label{SL}
\end{equation}%
together with a pair of boundary conditions. Here $(\sigma p_{rm})^{2}$ is a
separation constant. For sound speed profiles with a single minimum it is
well known (cf. e.g. Ref. 
\onlinecite{Brown-etal-95}
for the outline of a uniform asymptotic derivation) that an asymptotic
analysis of (\ref{SL}) for modes with turning depths within the water column
reveals that each $\phi _{m}$ is associated with a discrete value of the
action $I,$%
\begin{equation}
\sigma I(p_{rm})=m+\tfrac{1}{2},\quad m=0,1,2,\cdots ,  \label{m}
\end{equation}%
where $I(p_{r})$ is defined in (\ref{I}).

We now introduce, following the argument given in Ref. 
\onlinecite{Brown-etal-95}%
, the important concepts of modal \textit{group} and \textit{phase slowness}%
. Owing to the orthogonality of the modes, the quantity $\bar{u}%
_{m}(r,\sigma )=\int \mathrm{d}z$\thinspace $\bar{u}(z,r,\sigma )\phi
_{m}(z,\sigma )/\int \mathrm{d}z\,\phi _{m}^{2}(z,\sigma )$ isolates the
contribution to the wavefield from the mode with frequency $\sigma $ and
mode number $m.$ The inverse Fourier transformation of $\bar{u}_{m}(r,\sigma
)$, weighted by $\bar{s}(\sigma ),$ the Fourier transformation of the source
time history $s(t),$ is denoted $u_{m}(r,t).$ If $\bar{s}(\sigma )$ has a
narrow bandwidth, centered at $\sigma _{0},$ then a Taylor series expansion
of $k_{rm}=\sigma p_{rm}$, with $p_{rm}=p_{rm}(\sigma )$ via Eq. (\ref{m}),
about $\sigma _{0}$ yields the result%
\begin{equation}
u_{m}(r,t)=\mathrm{e}^{\mathrm{i}\left[ k_{rm}(\sigma _{0})r-\sigma _{0}t%
\right] }\Psi _{m}(r,t)  \label{WT}
\end{equation}%
where the envelope function $\Psi _{m}(r,t)$ travels at the group speed, $%
(\partial k_{r}/\partial \sigma )^{-1},$ evaluated at the center frequency
and mode number $m$. The group slowness is defined as%
\begin{equation}
S_{g}=\frac{\partial k_{r}}{\partial \sigma }.
\end{equation}%
Note that (\ref{WT}) represents a slowly varying dispersive wavetrain whose
envelope moves at the group slowness, but within which surfaces of constant
phase move at the phase slowness $k_{r}/\sigma =p_{r}.$

Consistent with the asymptotic analysis presented here%
\begin{equation}
S_{g}(p_{r})=\frac{T_{\ell }(p_{r})}{R_{\ell }(p_{r})}  \label{Sg}
\end{equation}%
(cf. e.g. Ref. 
\onlinecite{Munk-Wunsch-83}
or 
\onlinecite{Brown-etal-95}%
; the latter reference also includes, with additional references, the exact
expression for $S_{g}$). Here $R_{\ell }(p_{r})$ is given in (\ref{R}) and $%
T_{\ell }(p_{r})$ is the corresponding expression for the single-cycle
travel time,%
\begin{eqnarray}
T_{\ell }(p_{r}) &=&2\pi I(p_{r})+p_{r}R_{\ell }(p_{r})  \notag \\
&=&2\int_{z_{-}}^{z_{+}}\mathrm{d}z\,\frac{C^{-2}(z)}{\sqrt{%
C^{-2}(z)-p_{r}^{2}}}.
\end{eqnarray}%
Note that although, according to (\ref{Sg}), $S_{g}$ depends only on the
turning depths of a mode, this dictates via the quantization condition (\ref%
{m}) that $S_{g}$ is in general a function of both frequency $\sigma $ and
mode number $m.$

The waveguide invariant $\beta ,$ whose relationship to wavefield properties
will be reviewed in the next section, is defined as%
\begin{equation}
\beta =-\frac{\partial S_{g}}{\partial p_{r}}  \label{beta}
\end{equation}%
where $p_{r}$ is the modal phase slowness (often written as $S_{p}$). Like $%
S_{g}$, in general $\beta $ is a function of both $\sigma $ and $m.$ It
follows from Eqs. (\ref{R}, \ref{Sg}--\ref{beta}) that $\beta $ can be
expressed as%
\begin{equation}
\beta (p_{r})=2\pi \frac{I(p_{r})}{R_{\ell }^{2}(I)}\frac{\mathrm{d}R_{\ell }%
}{\mathrm{d}p_{r}}=\alpha (p_{r}).
\end{equation}%
This is the main result of this letter. It should be emphasized that our
modal analysis is based on asymptotic results, so that we have only
demonstrated the asymptotic equivalence of $\beta $ and $\alpha .$

\section{Wavefield structure and stability}

\begin{figure}[t]
\centerline{\includegraphics[width=8cm,clip=]{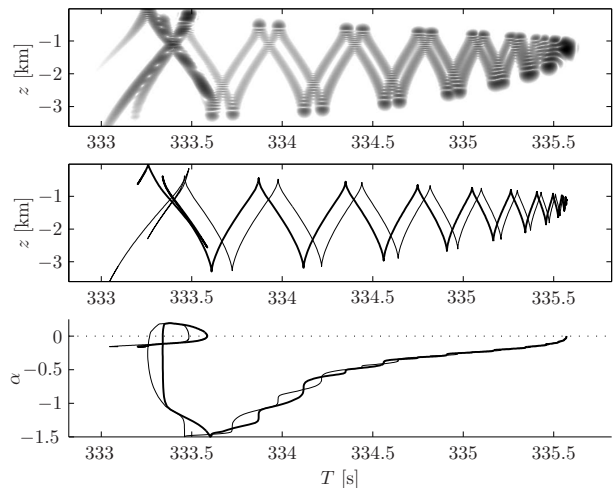}}
\caption{Upper panel: Intensity of a transient acoustic wavefield in a
range-independent deep-ocean environment as a function of depth and time at
a fixed range of $500$ \textrm{km}. The wavefield was excited by a compact
source near the sound channel axis with $f_{0}=75$ \textrm{Hz }and $\Delta
f\approx 30$ \textrm{Hz}. The dynamic range of intensity is $40$ \textrm{dB}%
. Middle panel: Corresponding ray travel times vs. depth. The light (heavy)
line corresponds to rays with an upward (downward) inclination at the
source. Lower panel: Corresponding plots of $\protect\alpha $ vs. $T$,
computed by eliminating the $I$-dependence from $\protect\alpha (I)$ and $%
T(I;r)$.}
\label{AlphaBeta}
\end{figure}

In this section we briefly review those properties of wavefield structure
and wavefield stability (to a small range-dependent perturbation) that are
known to be controlled by $\alpha $ or $\beta .$ Fig. \ref{AlphaBeta} shows
wavefield intensity $\left| u(z,r,t)\right| ^{2}$ in a range-independent
sound channel in the depth--time plane at a fixed range, $r=500$ \textrm{km,}
for waves excited by a transient compact source with center frequency $%
f_{0}=75$ $\mathrm{Hz}$ and bandwidth $\Delta f\approx 30$ $\mathrm{Hz}$
where $\sigma =2\pi f.$ This plot was produced by solving the
Thomson--Chapman \cite{Thomson-Chapman-83} parabolic equation. Use of a
parabolic approximation to the Helmholtz equation introduces some minor
distortion of the wavefield, but the connections described here between $%
\alpha ,$ $\beta ,$ wavefield structure and wavefield stability hold whether
a parabolic approximation is introduced or not.

Perhaps the simplest interpretation of $\alpha $ is that it is a measure of
the rate at which small elements of the extended phase space $(z,p,T)$ are
deformed by the background flow (with $\mathrm{d}z/\mathrm{d}r$ treated as
the $z$-coordinate of a fluid motion, etc.) by shearing motion. \cite%
{Beron-Brown-03a,Beron-Brown-03b} Although this property does not correspond
directly to any observable wavefield feature, this property helps to
understand other observable wavefield features. A very simple observable
wavefield property controlled by $\alpha $ is travel time dispersion. It
follows from the third of Eqs. (\ref{AAeqn}) that $\mathrm{d}T/\mathrm{d}I=I(%
\mathrm{d}\omega /\mathrm{d}I)r=\alpha (I)\omega (I)r.$ In a
range-independent environment $I$ is a ray label that increases
monotonically with increasing axial ray angle, so this equation describes
the rate of change of ray travel time with increasing axial ray angle. This
is illustrated in the ray simulation in Fig. \ref{AlphaBeta} where $\alpha
<0 $ except in small angular bands corresponding to steep rays; note that
this ray property is also clearly visible in the corresponding finite
frequency wavefield. Note also that zeros of $\alpha $ correspond
approximately to cusps in the $(z,T)$ plane. At such cusps geometric
amplitudes diverge. This can be seen from Eq. (\ref{a22}). In that
expression $R$ is the total range of a ray, which can be written as $%
nR_{\ell }$ ($n$ complete ray loops) plus end-segment corrections which
depend on source and receiver depths and ray inclination (positive or
negative) at the source and receiver. At long range the dominant
contribution to $R$ is $nR_{\ell }.$ Then it follows from (\ref{a22}) and (%
\ref{alpha2}) that at caustics either $p_{z}=0$ or $\alpha =0.$ [We
emphasize that is true only in the large $r$ asymptotic limit. At short
range $(\partial R/\partial p_{r})_{z}$ has complicated structure, usually
including a singularity when $p_{z}=0$, so that application of (\ref{a22})
generally requires great care near such points.] The same argument reveals
that at long range geometric amplitudes are inversely proportional to $%
\left| \alpha \right| ;$ this property is also seen (away from caustics and
interference fringes) in the finite frequency wavefield shown in Fig. \ref%
{AlphaBeta}.

In the presence of a weak range-dependent perturbation $\alpha $ has been
shown to control both ray stability, \cite{Beron-Brown-03a} quantified by
Lyapunov exponents, for instance, and several measures of travel time
spreads \cite{Beron-Brown-03b,Virovlyansky-03}. The latter property is
linked to the travel time dispersion property in the background environment
noted above. Also, in the presence of a weak perturbation, $\alpha $ plays a
critical role in ideas relating to ray dynamics. \cite%
{Brown-etal-03,Beron-etal-03,Smirnov-etal-01,Arnold-89,Tabor-89} References 
\onlinecite{Beron-Brown-03a,Beron-Brown-03b,Virovlyansky-03,Beron-etal-03}
show that unperturbed rays for which $\left| \alpha \right| $ is large are
generally more sensitive to environmental perturbations than those for which 
$\left| \alpha \right| $ is small.

In a range-independent environment $\beta $ controls modal group delay
dispersion: modal group delays statisfy $T_{g}=S_{g}r,$ where $%
T_{g}=T_{g}(m,\sigma )$ and $S_{g}=S_{g}(m,\sigma ),$ so $(\partial
T_{g}/\partial p_{r})_{r}=-\beta r.$ The property that is most commonly used %
\cite{Brekhovskikh-Lysanov-91} to motivate the introduction of the waveguide
invariant $\beta $ is that, following a surface of constant wavefield
intensity, a relationship $\delta r/r=\beta \delta \sigma /\sigma $ must be
satisfied. This leads naturally to the defining relationship (\ref{beta}).
(Note that in some references, including the original work by Chuprov, \cite%
{Chuprov-82} $-\partial S_{g}/\partial p_{r}$ is defined to be $1/\beta $
rather than $\beta .$) This defining relationship (\ref{beta}) implies a
power-law dependence of the difference in eigenvalues of the Helmholtz
equation on the acoustic frequency, $k_{rn}-k_{rm}=a_{mn}\sigma ^{-\beta }$
where $m$ and $n$ are mode numbers and $a_{mn}$ is a constant. More
generally Grachev \cite{Grachev-93} argued that $k_{rn}-k_{rm}=b_{mn}\eta
^{\beta /\gamma }\sigma ^{-\beta }$ where $\eta $ is a parameter, such as
water depth, describing the environment and $\gamma $ is a constant.
Consistent with this dependence is the relationship $\delta r/r=\beta \delta
\sigma /\sigma +(\beta /\gamma )\delta \eta /\eta .$ It is important to note
(cf. also Ref. 
\onlinecite{Kim-etal-03}%
) that because $\beta =\beta (\sigma ,m),$ when modes corresponding to
different values of $\beta $ are not temporally resolved, \cite%
{Brown-Viechnicki-Tappert-96} application of the above results may be
difficult. Much work relating to $\beta $ has focused on the homogeneous,
constant depth waveguide for which $\beta $ is independent of $\sigma $ and $%
m$ for modes corresponding to small values of axial ray angle; for that
problem $\beta $ can be approximated as a constant.

In an environment consisting of a range-indenepent background on which a
range-dependent perturbation is superimposed, $\beta $ controls the spread
of modal group delays owing to mode coupling. This is seen by noting that
the total group delay of modal energy that has been scattered among mode
numbers $m_{i},$ $i=1,\cdots ,N,$ is $T_{g}=\sum_{i}S_{g}(m_{i},\sigma
)r_{i} $ where $r=\sum_{i}r_{i}$ is the range. It follows that%
\begin{eqnarray}
T_{g} &=&S_{g}(\bar{m},\sigma )r+\frac{\partial S_{g}}{\partial p_{r}}(\bar{m%
},\sigma )\frac{\partial p_{r}}{\partial I}\sum_{i}(I_{i}-\bar{I})r_{i} 
\notag \\
&=&S_{g}(\bar{m},\sigma )r+\beta (\bar{m},\sigma )\frac{2\pi }{R_{\ell }(%
\bar{p}_{r})}\sum_{i}(I_{i}-\bar{I})r_{i}
\end{eqnarray}%
where $\bar{I}=(\bar{m}+\frac{1}{2})/\sigma =I(\bar{p}_{r})$ amd $\bar{m}$
is a suitably chosen mode number (e.g. the mode number at $r=0$ or the
average mode number). Also in range-dependent environments, the relationship 
$\delta r/r=\beta \delta \sigma /\sigma +(\beta /\gamma )\delta \eta /\eta $
has been used in the interpretation of measurements in shallow water
waveguides \cite{Weston-Stevens-72,Kuzkin-95,Kuzkin-etal-98,Kuzkin-98} and
time reversal applications. \cite%
{Kim-etal-03,Hodgkiss-etal-99,Dspain-etal-99,Song-etal-98} Note that large $%
\left| \beta \right| $ is associated with a high degree of wavefield
sensitivity, e.g. high sensitivity of modal group delays to an environmental
perturbation, or high sensitivity of the location of a time reversal focus
to an environmental perturbation $\delta \eta .$

In retrospect it is not surprising that there should be a simple connection
between $\alpha $ and $\beta $ inasmuch as, whether one adopts a ray or
modal wavefield description, the object of study, the wavefield, is the
same. The result $\beta \sim \alpha $ is an aspect of \textit{ray--mode
duality} (cf. e.g. Ref. 
\onlinecite{Munk-Wunsch-83}%
). We expect that there are many wavefield properties that we have
overlooked that are controlled by this parameter. We are unaware of any
parameter that characterizes an acoustic environment whose important rivals
that of $\alpha $ (or $\beta $).

\begin{acknowledgments}
This research was supported by Code 321OA of the Office of Naval Research.
\end{acknowledgments}

\end{document}